\documentclass[%
 reprint,
superscriptaddress,
nofootinbib,
 amsmath,amssymb,
 aps,
floatfix
]{revtex4-2}

\usepackage{graphicx}
\usepackage{dcolumn}
\usepackage{bm}

\usepackage{siunitx}
\usepackage{hyperref}
\usepackage{ragged2e}
\usepackage{float}

\usepackage{xcolor}
\usepackage{soul}
\usepackage[normalem]{ulem}
\usepackage{subcaption}
\usepackage{chngcntr}
\usepackage{mwe}

\begin{document}

\preprint{APS/123-QED}

\title{JDAM  - Jump Diffusion by Analytic Models}

\author{Yaqing Xy Wang}%
\affiliation{Cavendish Laboratory, University of Cambridge, 19 J J Thomson Avenue, Cambridge CB3 0HE, United Kingdom}

\author{Jack Kelsall}%
\affiliation{Cavendish Laboratory, University of Cambridge, 19 J J Thomson Avenue, Cambridge CB3 0HE, United Kingdom}

\author{Nadav Avidor}%
\email{na364@cam.ac.uk}
\affiliation{Cavendish Laboratory, University of Cambridge, 19 J J Thomson Avenue, Cambridge CB3 0HE, United Kingdom}

\date{\today}

\begin{abstract}
Nanoscopic diffusion at surfaces normally takes place when an adsorbate jumps from one adsorption site to the other. Jump diffusion can be measured via quasi-elastic scattering experiments, and the results can often be interpreted in terms of analytic models. While the simplest model of jump diffusion only accounts for intercell jumps between nearest neighbours, recent works have highlighted that models which take into account both intracell and long-range intercell jumps are much needed. Here, we describe a program to compute the analytic lineshape expected from quasi-elastic scattering experiments, for translational jump diffusion on user-defined lattices. We provide an example of a general hexagonal surface composed of six sublattices, corresponding to the six principle adsorption sites - namely the top, two hollow, and three bridge sites. In that example we include only nearest-neighbour jumps. In addition, we provide a mean to calculate the lineshape for jumps on a hexagonal honeycomb lattice, with jumps up to the $10^{\mathrm{th}}$ nearest neighbour taken into consideration.

\end{abstract}

\maketitle

\section{PROGRAM SUMMARY}

\begin{table*}[!t]
\begin{tabular}{|l|l|}
\hline
\textbf{Code metadata description} & \\
\hline
Current code version & v1.1 \\
\hline
Github repository &  $https://github.com/na364/JDAM$ \\
\hline
Permanent link to code/repository used of this code version &  $https://doi.org/10.5281/zenodo.4049976$ \\
\hline
Legal Code License   & GNU/GPL-3.0 \\
\hline
Code versioning system used & git \\
\hline
Software code languages, tools, and services used & MATLAB \\
\hline
Support email for questions & atomscattering@phy.cam.ac.uk \\
\hline
\end{tabular}
\caption{Current code version}
\label{curr_code_ver}
\end{table*}

Nature of problem: Molecular diffusion on hexagonal surfaces can be measured using quasi-elastic scattering techniques. The analysis of such measurements often requires the experimental data to be interpreted in terms of analytic models for translational jump diffusion on non-Bravais lattices. To date, a few models have been described analytically in the literature. However, those cases only consider high-symmetry azimuths of hexagonal surfaces and a single type of adsorption site; more specifically, only jumps on threefold hollow sites and on twofold bridge sites have been studied. However, for many adsorbate-substrate systems, azimuths without high symmetry and multiple types of adsorption site are relevant and must be incorporated into models in order to explain the observed surface dynamics.

Solution method: We provide a program which solves the general case of jump diffusion on a lattice. The code calculates the lineshape of the intermediate scattering function (ISF), the observable in quasi-elastic scattering experiments. To assist the user, we provide an example in which the ISF is calculated for jump diffusion on the six principle adsorption sites of a hexagonal surface, including only nearest-neighbour jumps. The code can be modified to included longer jumps and additional types of lattices. We also include the functionality to compute the ISF for the case of jumps between hexagonal hollow sites, for up to the $10^{\mathrm{th}}$ nearest neighbour.

\section{Introduction}

The transport of atoms and molecules at surfaces is a subject of key scientific importance, underpinning processes in areas such as catalysis, environmental science, and interstellar chemistry  \cite{riscoe2019transition,brasseur2017modeling,lamberts2016quantum}. The emergence of quasi-elastic scattering techniques, in particular of helium and neutron spin-echo spectroscopy, allows the study of atomic and molecular dynamics at elevated temperatures and on picosecond time scales \cite{alexandrowicz2007helium,jardine2009helium}. Helium spin-echo spectroscopy (HeSE) was first utilized to study the bound-state resonances of scattered helium atoms at the surface \cite{jardine2004ultrahigh,riley2007refined,avidor2016helium}. It has subsequently been applied to study jump diffusion of particles between adsorption wells \cite{tamtogl2020nanoscopic}. In contrast to techniques that measure at low temperatures, in which particles strongly prefer to occupy the lowest-energy adsorption sites, HeSE allows the population of particles in higher-energy wells to be resolved. HeSE is therefore able to provide a complete picture of the energy landscapes of the surface.

When analyzing quasi-elastic scattering data for surface adsorbates undergoing jump diffusion, an analytic model can serve as a useful reference when fitting the data and in the interpretation of the diffusion mechanism. Several examples of such models already exist, the simplest being the Chudley-Elliott (CE) model \cite{chudley1961neutron}. The CE model describes intercell jump diffusion on a Bravais lattice. The model has been extensively used over the years. However, with the emergence of helium spin-echo spectroscopy which is able to measure intracell jump diffusion, extensions to the model have been sought.

Tuddenham \textit{et.al.}\cite{tuddenham2010lineshapes} have presented a general method for modeling single jumps between non-equivalent surface sites, \textit{i.e.} diffusion on a lattice with non-trivial basis. The authors provided explicit analytic expressions for the ISF dephasing rates and the corresponding intensities for jumps on a basis composed of hollow sites, and for jumps on bridge sites. In both cases, only the high symmetry azimuths of the lattice were considered.

Further to Tuddenham \textit{et.al.}, Townsend and Avidor\cite{townsend2019signatures} have extended the hollow-site diffusion model to include up to fourth-order jumps. They have provided explicit expressions that allow calculation of the ISF in terms of just three parameters. Again, only the high symmetry azimuths are considered. However, it is expected that not all adsorbate-substrate systems adhere to hollow site or bridge site jump diffusion models. As demonstrated by Kole \textit{et.al.} in a study of CO on Cu(111) \cite{kole2012probing}, there are systems in which adsorbates jump between the top and bridge sites. Such cases cannot be modeled by the aforementioned published analytic models. Inevitably, it is expected that other paradigms for jump diffusion exist, especially when there are multiple adsorption sites in the unit cell. Furthermore, when studying the diffusion of molecules with a rotational symmetry different to that of the substrate, it is important to be able to evaluate ISFs along additional azimuths other than those with high symmetry.

In this work we utilize the general framework for diffusion on lattice with basis and present a software code to model the general case of jump diffusion on all the principle adsorption sites (top, bridge, and hollow) of a hexagonal surface. We name the software JDAM - `Jump Diffusion by Analytic Models'. In addition, we provide a script to evaluate the ISF for jumps to higher order neighbour sites on a honeycomb lattice, on all crystal azimuths. Overall, we aim to open up the way for a more accurate analysis of past and future surface spin-echo measurements.

\section{Theory}

We provide a summary of the theory, details of which were first outlined for surfaces by Tuddenham \textit{et.al.} \cite{tuddenham2010lineshapes}.
Quasi-elastic scattering measurements in the time domain record the ISF, $\textit{I}(\textbf{Q},t)$, which is a spatial Fourier transform of the Van Hover pair correlation function $\textit{G}(\textbf{r},t)$. Classically, $\textit{G}(\textbf{r},t)$ can be interpreted as the probability of finding a particle at position $(\textbf{r},t)$, given that there was a particle at $(\textbf{r}=0,t=0)$.  As such, the ISF contains a complete picture of the microscopic dynamics at the surface. For a static system, the ISF is a constant function of unity because there exists no decay of correlation over time. However, for a dynamical system, there will be a loss of surface correlation with increasing time, causing the ISF to decay. For jump diffusion on a lattice, the ISF is a sum of exponential decays.

The adsorption lattice we consider is a Bravais lattice consisting of adsorption sites which form $m$ sublattices. In the single jump regime, a particle can jump from a sub-lattice $i$ to another $j$ via $n_{ij}$ routes, with a jump vector $\textbf{l}=\textbf{L}_{ijk}$ with the index $k$ running from $1$ to $n_{ij}$. The probability $\textit{P}_i(\textbf{r},t)$ of a particle being at position $\textbf{r}$ in the $\textit{i}$th sub-lattice at time t is

$$\frac{\partial \textit{P}_i(\textbf{r},t)}{\partial t} = \sum_{j,k}\left[\frac{ \textit{P}_j(\textbf{r}+\textbf{L}_{ijk})}{\tau_{jik}}-\frac{ \textit{P}_i(\textbf{r},t)}{ \tau_{ijk}}\right],$$

\noindent
where $\tau_{ijk}^{-1}$ is the jump rate for a jump with $\textbf{l}=\textbf{L}_{ijk}$. After a spatial Fourier transform and collecting the different components of the ISF into a vector $\mathbf{I}$, we obtain the ISF, 

$$\frac{\partial}{\partial t}\textbf{\textit{I}}(\textbf{Q},t) = \textbf{A\textit{I}}(\textbf{Q},t),$$

\noindent
where $\textbf{A}$ is a matrix with elements 

$$\textit{A}_{ij} = \sum_{k}\frac{1}{\tau_{jik}}\exp(-i\textbf{Q}\cdot\textbf{L}_{ijk})-\delta_{ij}\sum_{j'}\frac{1}{\tau_{ij'}}.$$

\noindent
We then convert the matrix to a Hermitian matrix,

$$\textbf{B} = \textbf{TAT}^{-1},$$

\noindent
using the similarity transformation $\mathbf{T}$,

$$T_{ij} = \sqrt{\frac{1}{c_i}}\delta_{ij}.$$

\noindent
Equivalently, 

$$\textit{I}(\textbf{Q},t)= \sum_{p}w_{p}(\textbf{Q})\exp\left[M_{p}(\textbf{Q})t\right],$$

\noindent
where $M_{p}$ is the $p$th eigenvalue of the $\textbf{B}$ matrix,equivalently the dephasing rates or decay constants and 

$$w_{p} (\textbf{Q}) = \left|\sum_{i}\sqrt{c_{i}}b_{i}^{p}\right|^{2}$$ 

\noindent
where $b_{i}^{p}$ denotes the $i$th element of the $p$th eigenvector of $\textbf{B}$ and  $c_{i}$ is the adsorbate concentration on site $i$.

\section{Package overview}

JDAM is coded in MATLAB, and calculates the ISF for surfaces and parameters defined by the user. As an example, we have provided a file defining a hexagonal surface with the code, for jumps to nearest neighbours. We have also included a code which calculates jumps to sites beyond the nearest neighbours for the case of diffusion on honeycomb lattice. The code is readily expandable both for other types of surfaces, and to jumps of higher order. We here briefly describe the user parameters, the structure of the program, and the hexagonal surface. 

\subsection{Package Structure}
We here describe the main functions in JDAM.
\begin{enumerate}
    \item hex\_ui.m: user interface file. Contains the system parameters that may be varied by the user. The user can alter thermodynamic variables such as temperature, residence time on each sub-lattice, and adsorption energies on each of the sublattices. In addition, the file specifies the type of surface and the azimuths along which the lineshape is calculated. The user may also set the model to test modes that exclusively compute ISFs for single-site jump diffusion; specifically, for diffusion on top sites, bridge sites and hollow sites.
    \item surf\_gen.m: A function for generating the surface lattice using predefined lattice vectors.
    \item calc\_ISF.m: Here, the $\textbf{A}$ matrix is computed and transformed to the Hermitian matrix $\textbf{B}$. Its eigenvalues and eigenvectors are subsequently calculated. The complete lineshape is then calculated as the weighted sum explained before.
    \item hex\_vectors.m: A file that contains the relevant jump vectors. For each type of lattice, a different file should be provided.
    \item run\_jdam.m: The root file that constructs the lattice from user configurations and then plots the ISFs (dephasing rates and intensities) after calculating them.
\end{enumerate}

\subsection{User inputs}
\begin{enumerate}
\item In this software presented, the user may supply \textbf{
$\tau$} matrix as input, where $\tau_{ij}^{-1}$ is interpreted as the jump frequency from site $i$ to site $j$. Physically this quantity is subject to the Arrhenius Form:

$$\frac{1}{\tau}=\Gamma_{0}\exp\left(-\frac{E}{k_BT}\right),$$

\nonumber
where $\Gamma_{0}$ is the attempt frequency, $E$ is the activation energy (energy barrier) along the jump, $k_B$ is the Boltzmann constant, and $T$ is the system temperature. The expression may be useful as a reference for an \textit{ab initio} estimate of the jump rate. The user is required to configure the upper triangular part of the matrix as the lower triangular part will be calculated based on site degeneracy or non-degeneracy.
\item The site energy profile, corresponding to values of $\epsilon_{i}$, is then configured. $\epsilon_{i}$ is the energy minimum of the potential well at each site. Thermodynamically, the probability that a particle resides on a site with energy $\epsilon$ is proportional to

$$\frac{1}{Z}\exp\left(-\frac{\epsilon}{k_BT}\right),$$ 

\noindent
where $Z$ is the partition function of the system. As a result, the ratio of adsorbate concentrations for two sites with energy minima $\epsilon_{i}$ and $\epsilon_{j}$, $c_{i}/c_{j}$, is given by

$$\lambda_{ij}=\exp\left(\frac{\epsilon_{j}-\epsilon_{i}}{k_BT}\right).$$

\item The site degeneracy is then a property consistent with site concentrations. In the steady state, two sites with the same energy minima will have the same adsorbate concentration, i.e. $\lambda = 1$. In such a steady state the rate of outgoing jumps $i$ to $j$ must be the same as that of incoming jumps $j$ to $i$, i.e.

$$c_{i}\tau_{ij}^{-1}=c_{j}\tau_{ji}^{-1}.$$ 

\noindent
The above is how the lower triangular part of the \textbf{$\tau$} is calculated.
\end{enumerate}

\subsection{Lattice}

\begin{figure} [h!]
\centering
    \includegraphics[width=0.4\textwidth]{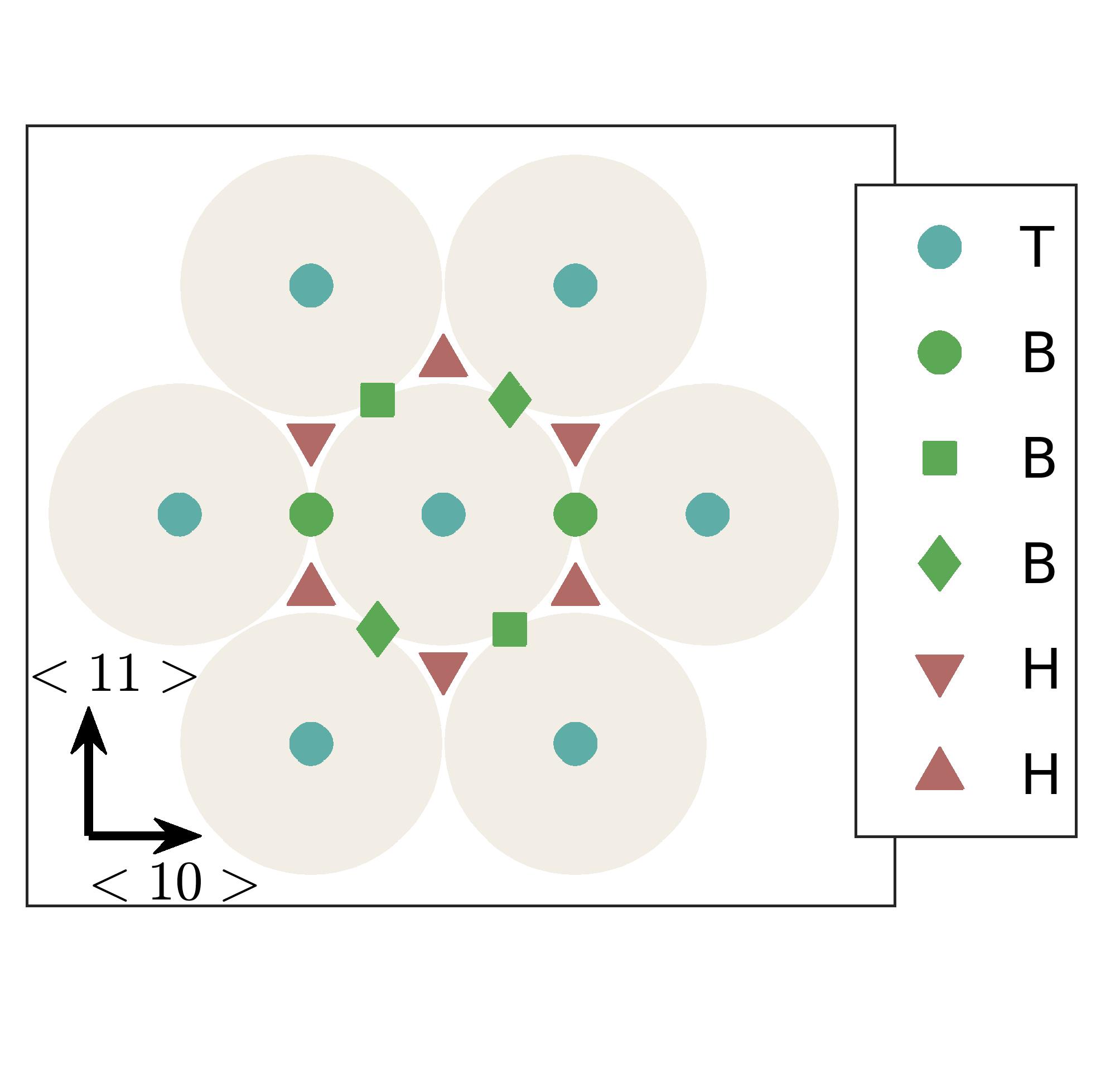}
    \caption{\label{fig:hex_lattice} An illustration of a hexagonal surface. The six types of sublattices incorporated into JDAM are marked. The distinct lattices are a lattice of top sites (T), three sublattices of bridge sites (B), and two sublattices of hollow sites (H) which are normally referred to as the HCP and FCC adsorption sites. The script supplied in JDAM considers jumps to nearest neighbours. For example, a particle which is positioned at the central top site would be allowed to jump to all the sites which are sketched in the figure. In addition, JDAM allows for the special case of jumps up to $10^{\mathrm{th}}$ order neighbours on hollow sites, when only hollow sites are considered.}
\end{figure}

The JDAM package includes an example file which defines the jump vectors on a hexagonal surface. The surface adsorption sites form a Bravais lattice with six sublattices; the unit cell is illustrated in Figure \ref{fig:hex_lattice}. As shown, the six sites include one top site, three twofold bridge sites (degenerate in site energy) and two threefold hollow sites. The jump vectors $\textbf{L}_{ijk}$ and the matrix $\mathbf{n}$, which encodes the number of distinct routes a jump from site $i$ to site $j$ can take, are both calculated directly from the configuration of the surface lattice which is illustrated.

\section{Benchmarking}

\begin{figure*}
    \centering
    \begin{subfigure}{0.333\textwidth}
        \centering
        \includegraphics[width=1.0\textwidth]{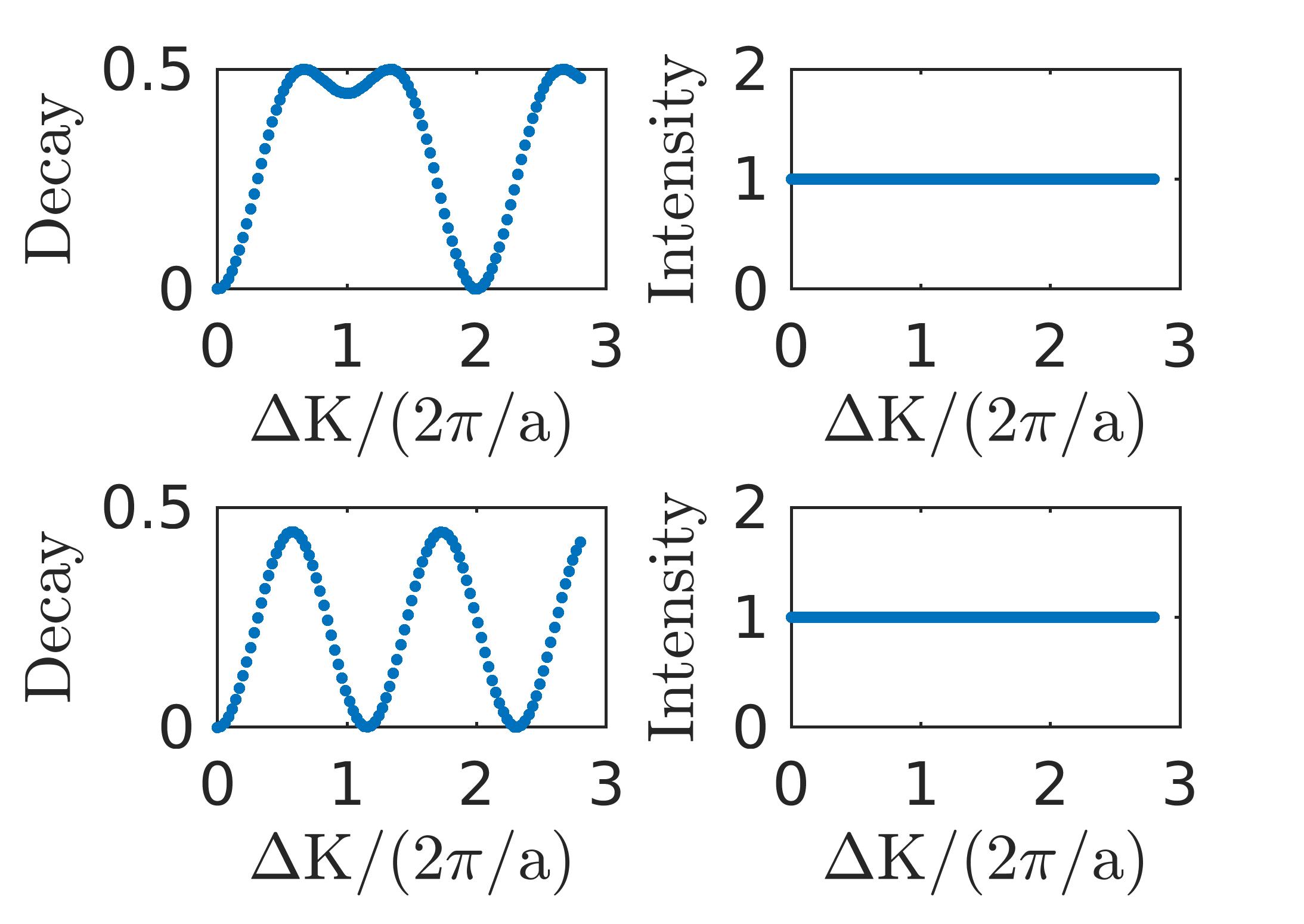}
        \caption{\label{benchtop}}
    \end{subfigure}%
    \begin{subfigure}{0.333\textwidth}
        \centering
        \includegraphics[width=1.0\textwidth]{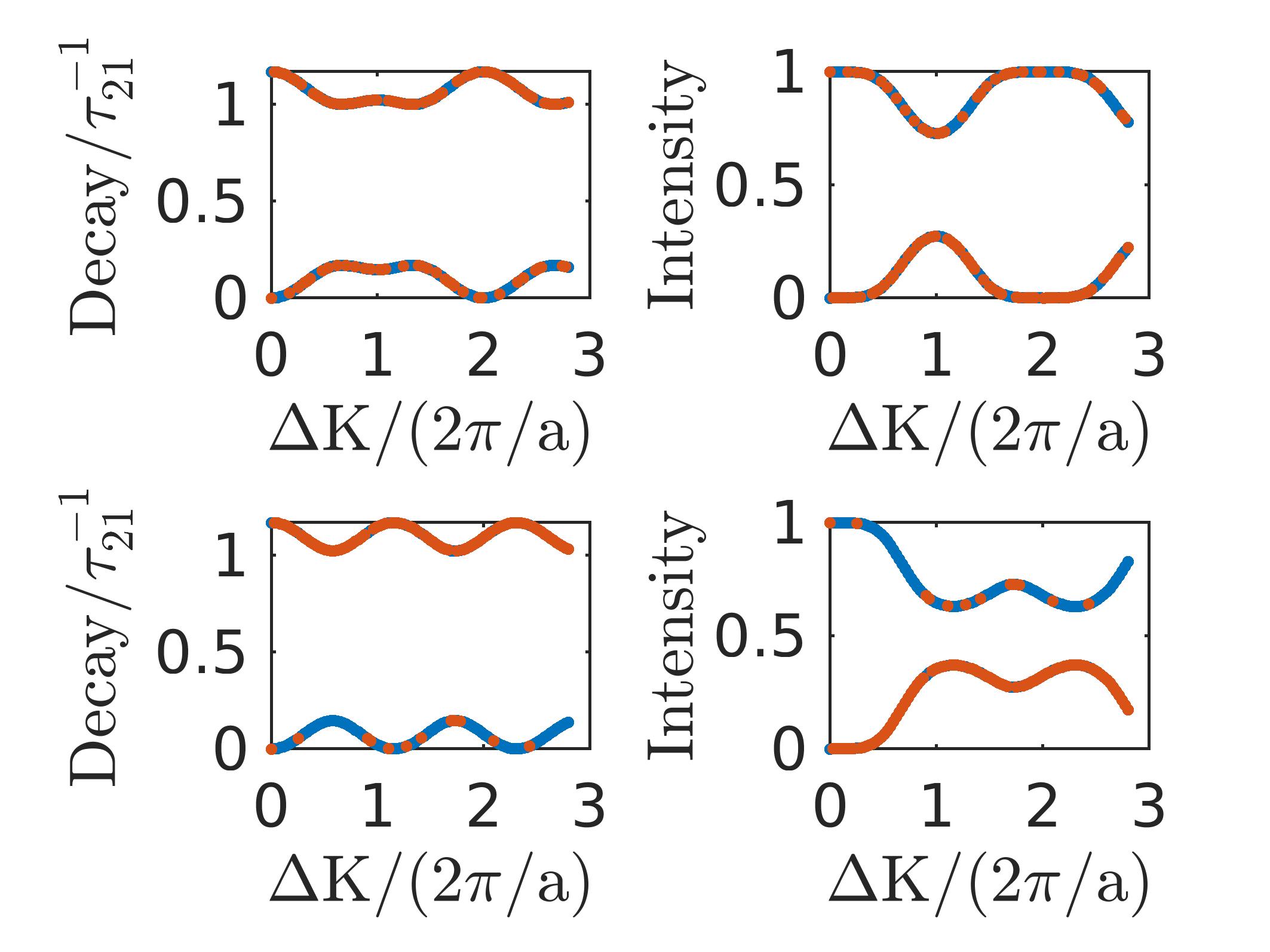}
        \caption{\label{benchhollow}}
    \end{subfigure}%
    \begin{subfigure}{0.333\textwidth}
        \centering
        \includegraphics[width=1.0\textwidth]{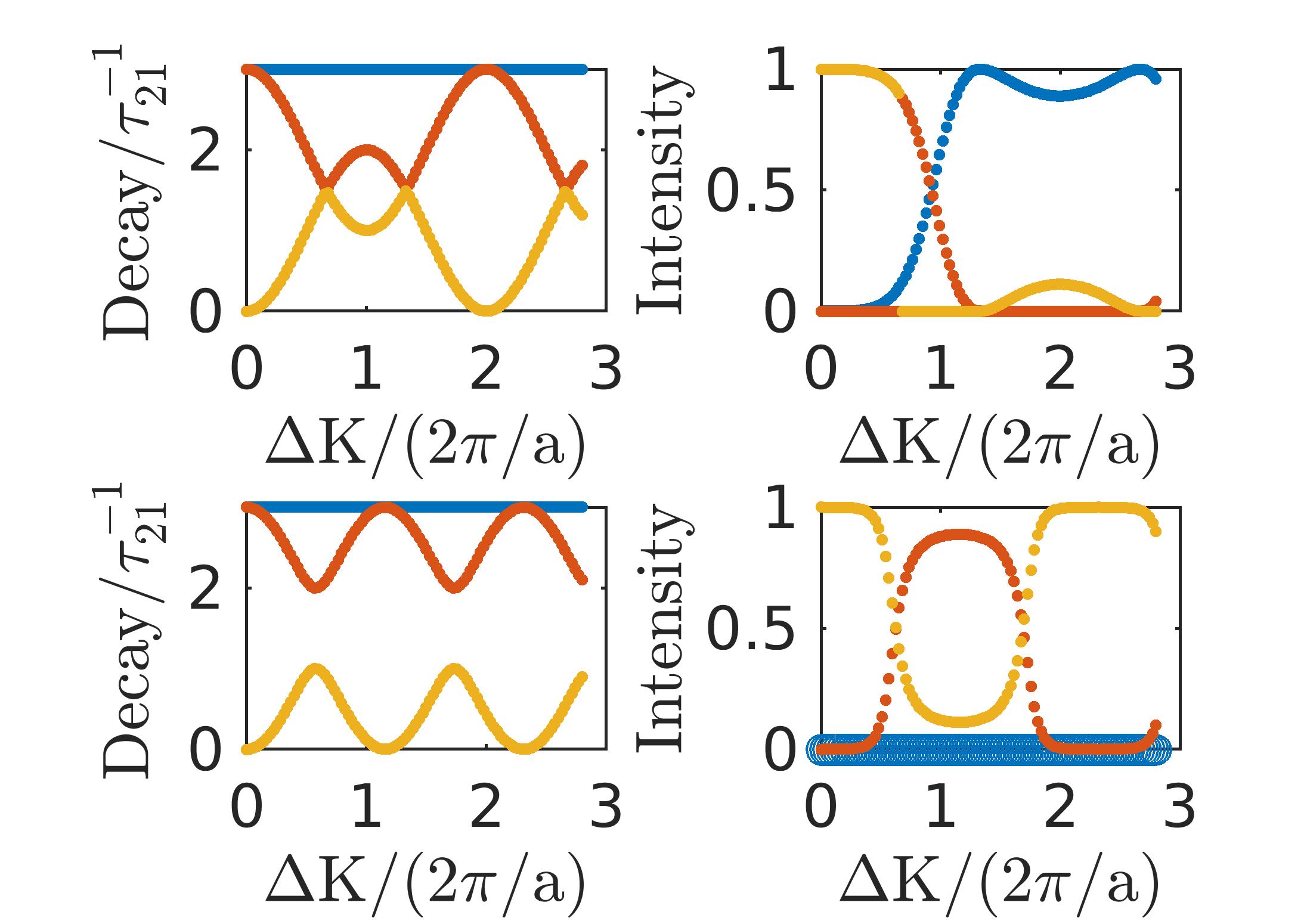}
        \caption{\label{benchbridge}}
    \end{subfigure}
    \caption{\label{benchmarks} Results from JDAM compared with explicit expressions for a simple Chudley-Elliot model, and for hopping on a non-Bravais lattice as given by Tuddenham \textit{et.al.} \cite{tuddenham2010lineshapes}. For each subfigure, the upper panel displays the results for the $[11\bar{2}]$ surface azimuth, and the lower panel for the $[1\bar{1}0]$ azimuth. (a) Dephasing rate scaled by $\tau_{21}^{-1}$  and normalized intensities for single jumps on top sites, $\lambda=1$. (b) Dephasing rate scaled by $\tau_{21}^{-1}$ and normalized intensities for single jumps on threefold hollow sites where the HCP potential well minimum is set to $0$ meV and the FCC minimum is $46$ meV, yielding $\lambda=5.911$. (c) Dephasing rate scaled by $\tau_{21}^{-1}$  and normalized intensities of single jumps on twofold bridge sites where all three site energies are degenerate, meaning $\lambda=1$}
\end{figure*}

The code in the present paper has been tested against the cases considered by Tuddenham \textit{et al.} for benchmarking purposes \cite{tuddenham2010lineshapes}. See Figure \ref{benchmarks} for:
\begin{enumerate}
    \item single jumps on threefold hollow sites, for $\lambda=5.9$, and
    \item single jumps on twofold bridge sites with degenerate site energies and concentrations.
\end{enumerate}
 
\section{Summary}
We present a package, JDAM, for computing the ISF of jump diffusion on a general non-Bravais lattice. JDAM includes an explicit example for calculating the ISF for the case of jumps on a hexagonal close-packed surface. In the example, all the principle adsorption sites are taken into account. Namely, all six sublattices are included, corresponding to adsorption on top, three bridge, and two hollow sites in the unit cell. Currently, only jumps to nearest neighbours are considered in the general case. In addition, we include an explicit example for the calculation of the ISF for jumps on hollow sites up to $10^{\mathrm{th}}$ nearest neighbours.

\begin{acknowledgments}
The work was conducted as part of the Undergraduate Summer Research Program, `Surface nano Physics and Atom-Surface Scattering', provided by the Cambridge Atom Scatting Facility (CASF) at the Cavendish Laboratory, University of Cambridge. Y. W. gratefully acknowledges CASF, the Cavendish Surface Physics group, and the group of supervisors and speakers for the opportunity to participate and conduct research during the program. The Engineering and Physical Sciences Research Council (EPSRC) is acknowledged for the financial support to CASF (EP/T00634X/1). N. A. gratefully acknowledges the Herchel Smith for funding. J. K. gratefully acknowledges the EPSRC for doctoral funding.
\end{acknowledgments}

\bibliography{jdam}

\begin{thebibliography}{13}%
\makeatletter
\providecommand \@ifxundefined [1]{%
 \@ifx{#1\undefined}
}%
\providecommand \@ifnum [1]{%
 \ifnum #1\expandafter \@firstoftwo
 \else \expandafter \@secondoftwo
 \fi
}%
\providecommand \@ifx [1]{%
 \ifx #1\expandafter \@firstoftwo
 \else \expandafter \@secondoftwo
 \fi
}%
\providecommand \natexlab [1]{#1}%
\providecommand \enquote  [1]{``#1''}%
\providecommand \bibnamefont  [1]{#1}%
\providecommand \bibfnamefont [1]{#1}%
\providecommand \citenamefont [1]{#1}%
\providecommand \href@noop [0]{\@secondoftwo}%
\providecommand \href [0]{\begingroup \@sanitize@url \@href}%
\providecommand \@href[1]{\@@startlink{#1}\@@href}%
\providecommand \@@href[1]{\endgroup#1\@@endlink}%
\providecommand \@sanitize@url [0]{\catcode `\\12\catcode `\$12\catcode
  `\&12\catcode `\#12\catcode `\^12\catcode `\_12\catcode `\%12\relax}%
\providecommand \@@startlink[1]{}%
\providecommand \@@endlink[0]{}%
\providecommand \url  [0]{\begingroup\@sanitize@url \@url }%
\providecommand \@url [1]{\endgroup\@href {#1}{\urlprefix }}%
\providecommand \urlprefix  [0]{URL }%
\providecommand \Eprint [0]{\href }%
\providecommand \doibase [0]{https://doi.org/}%
\providecommand \selectlanguage [0]{\@gobble}%
\providecommand \bibinfo  [0]{\@secondoftwo}%
\providecommand \bibfield  [0]{\@secondoftwo}%
\providecommand \translation [1]{[#1]}%
\providecommand \BibitemOpen [0]{}%
\providecommand \bibitemStop [0]{}%
\providecommand \bibitemNoStop [0]{.\EOS\space}%
\providecommand \EOS [0]{\spacefactor3000\relax}%
\providecommand \BibitemShut  [1]{\csname bibitem#1\endcsname}%
\let\auto@bib@innerbib\@empty
\bibitem [{\citenamefont {Riscoe}\ \emph {et~al.}(2019)\citenamefont {Riscoe},
  \citenamefont {Wrasman}, \citenamefont {Herzing}, \citenamefont {Hoffman},
  \citenamefont {Menon}, \citenamefont {Boubnov}, \citenamefont {Vargas},
  \citenamefont {Bare},\ and\ \citenamefont
  {Cargnello}}]{riscoe2019transition}%
  \BibitemOpen
  \bibfield  {author} {\bibinfo {author} {\bibfnamefont {A.~R.}\ \bibnamefont
  {Riscoe}}, \bibinfo {author} {\bibfnamefont {C.~J.}\ \bibnamefont {Wrasman}},
  \bibinfo {author} {\bibfnamefont {A.~A.}\ \bibnamefont {Herzing}}, \bibinfo
  {author} {\bibfnamefont {A.~S.}\ \bibnamefont {Hoffman}}, \bibinfo {author}
  {\bibfnamefont {A.}~\bibnamefont {Menon}}, \bibinfo {author} {\bibfnamefont
  {A.}~\bibnamefont {Boubnov}}, \bibinfo {author} {\bibfnamefont
  {M.}~\bibnamefont {Vargas}}, \bibinfo {author} {\bibfnamefont {S.~R.}\
  \bibnamefont {Bare}},\ and\ \bibinfo {author} {\bibfnamefont
  {M.}~\bibnamefont {Cargnello}},\ }\bibfield  {title} {\bibinfo {title}
  {Transition state and product diffusion control by polymer--nanocrystal
  hybrid catalysts},\ }\href@noop {} {\bibfield  {journal} {\bibinfo  {journal}
  {Nature Catalysis}\ }\textbf {\bibinfo {volume} {2}},\ \bibinfo {pages} {852}
  (\bibinfo {year} {2019})}\BibitemShut {NoStop}%
\bibitem [{\citenamefont {Brasseur}\ and\ \citenamefont
  {Jacob}(2017)}]{brasseur2017modeling}%
  \BibitemOpen
  \bibfield  {author} {\bibinfo {author} {\bibfnamefont {G.~P.}\ \bibnamefont
  {Brasseur}}\ and\ \bibinfo {author} {\bibfnamefont {D.~J.}\ \bibnamefont
  {Jacob}},\ }\href@noop {} {\emph {\bibinfo {title} {Modeling of atmospheric
  chemistry}}}\ (\bibinfo  {publisher} {Cambridge University Press},\ \bibinfo
  {year} {2017})\BibitemShut {NoStop}%
\bibitem [{\citenamefont {Lamberts}\ \emph {et~al.}(2016)\citenamefont
  {Lamberts}, \citenamefont {Samanta}, \citenamefont {K{\"o}hn},\ and\
  \citenamefont {K{\"a}stner}}]{lamberts2016quantum}%
  \BibitemOpen
  \bibfield  {author} {\bibinfo {author} {\bibfnamefont {T.}~\bibnamefont
  {Lamberts}}, \bibinfo {author} {\bibfnamefont {P.~K.}\ \bibnamefont
  {Samanta}}, \bibinfo {author} {\bibfnamefont {A.}~\bibnamefont {K{\"o}hn}},\
  and\ \bibinfo {author} {\bibfnamefont {J.}~\bibnamefont {K{\"a}stner}},\
  }\bibfield  {title} {\bibinfo {title} {Quantum tunneling during interstellar
  surface-catalyzed formation of water: the reaction {$\textrm{H} +
  \textrm{H}_2\textrm{O}_2 \longrightarrow \textrm{H}_2\textrm{O} +
  \textrm{OH}$}},\ }\href@noop {} {\bibfield  {journal} {\bibinfo  {journal}
  {Physical Chemistry Chemical Physics}\ }\textbf {\bibinfo {volume} {18}},\
  \bibinfo {pages} {33021} (\bibinfo {year} {2016})}\BibitemShut {NoStop}%
\bibitem [{\citenamefont {Alexandrowicz}\ and\ \citenamefont
  {Jardine}(2007)}]{alexandrowicz2007helium}%
  \BibitemOpen
  \bibfield  {author} {\bibinfo {author} {\bibfnamefont {G.}~\bibnamefont
  {Alexandrowicz}}\ and\ \bibinfo {author} {\bibfnamefont {A.}~\bibnamefont
  {Jardine}},\ }\bibfield  {title} {\bibinfo {title} {Helium spin-echo
  spectroscopy: studying surface dynamics with ultra-high-energy resolution},\
  }\href@noop {} {\bibfield  {journal} {\bibinfo  {journal} {Journal of
  Physics: Condensed Matter}\ }\textbf {\bibinfo {volume} {19}},\ \bibinfo
  {pages} {305001} (\bibinfo {year} {2007})}\BibitemShut {NoStop}%
\bibitem [{\citenamefont {Jardine}\ \emph {et~al.}(2009)\citenamefont
  {Jardine}, \citenamefont {Hedgeland}, \citenamefont {Alexandrowicz},
  \citenamefont {Allison},\ and\ \citenamefont {Ellis}}]{jardine2009helium}%
  \BibitemOpen
  \bibfield  {author} {\bibinfo {author} {\bibfnamefont {A.}~\bibnamefont
  {Jardine}}, \bibinfo {author} {\bibfnamefont {H.}~\bibnamefont {Hedgeland}},
  \bibinfo {author} {\bibfnamefont {G.}~\bibnamefont {Alexandrowicz}}, \bibinfo
  {author} {\bibfnamefont {W.}~\bibnamefont {Allison}},\ and\ \bibinfo {author}
  {\bibfnamefont {J.}~\bibnamefont {Ellis}},\ }\bibfield  {title} {\bibinfo
  {title} {Helium-3 spin-echo: Principles and application to dynamics at
  surfaces},\ }\href@noop {} {\bibfield  {journal} {\bibinfo  {journal}
  {Progress in Surface Science}\ }\textbf {\bibinfo {volume} {84}},\ \bibinfo
  {pages} {323} (\bibinfo {year} {2009})}\BibitemShut {NoStop}%
\bibitem [{\citenamefont {Jardine}\ \emph {et~al.}(2004)\citenamefont
  {Jardine}, \citenamefont {Dworski}, \citenamefont {Fouquet}, \citenamefont
  {Alexandrowicz}, \citenamefont {Riley}, \citenamefont {Lee}, \citenamefont
  {Ellis},\ and\ \citenamefont {Allison}}]{jardine2004ultrahigh}%
  \BibitemOpen
  \bibfield  {author} {\bibinfo {author} {\bibfnamefont {A.~P.}\ \bibnamefont
  {Jardine}}, \bibinfo {author} {\bibfnamefont {S.}~\bibnamefont {Dworski}},
  \bibinfo {author} {\bibfnamefont {P.}~\bibnamefont {Fouquet}}, \bibinfo
  {author} {\bibfnamefont {G.}~\bibnamefont {Alexandrowicz}}, \bibinfo {author}
  {\bibfnamefont {D.~J.}\ \bibnamefont {Riley}}, \bibinfo {author}
  {\bibfnamefont {G.~Y.}\ \bibnamefont {Lee}}, \bibinfo {author} {\bibfnamefont
  {J.}~\bibnamefont {Ellis}},\ and\ \bibinfo {author} {\bibfnamefont
  {W.}~\bibnamefont {Allison}},\ }\bibfield  {title} {\bibinfo {title}
  {Ultrahigh-resolution spin-echo measurement of surface potential energy
  landscapes},\ }\href@noop {} {\bibfield  {journal} {\bibinfo  {journal}
  {Science}\ }\textbf {\bibinfo {volume} {304}},\ \bibinfo {pages} {1790}
  (\bibinfo {year} {2004})}\BibitemShut {NoStop}%
\bibitem [{\citenamefont {Riley}\ \emph {et~al.}(2007)\citenamefont {Riley},
  \citenamefont {Jardine}, \citenamefont {Dworski}, \citenamefont
  {Alexandrowicz}, \citenamefont {Fouquet}, \citenamefont {Ellis},\ and\
  \citenamefont {Allison}}]{riley2007refined}%
  \BibitemOpen
  \bibfield  {author} {\bibinfo {author} {\bibfnamefont {D.}~\bibnamefont
  {Riley}}, \bibinfo {author} {\bibfnamefont {A.}~\bibnamefont {Jardine}},
  \bibinfo {author} {\bibfnamefont {S.}~\bibnamefont {Dworski}}, \bibinfo
  {author} {\bibfnamefont {G.}~\bibnamefont {Alexandrowicz}}, \bibinfo {author}
  {\bibfnamefont {P.}~\bibnamefont {Fouquet}}, \bibinfo {author} {\bibfnamefont
  {J.}~\bibnamefont {Ellis}},\ and\ \bibinfo {author} {\bibfnamefont
  {W.}~\bibnamefont {Allison}},\ }\bibfield  {title} {\bibinfo {title} {A
  refined he--lif (001) potential from selective adsorption resonances measured
  with high-resolution helium spin-echo spectroscopy},\ }\href@noop {}
  {\bibfield  {journal} {\bibinfo  {journal} {The Journal of chemical physics}\
  }\textbf {\bibinfo {volume} {126}},\ \bibinfo {pages} {104702} (\bibinfo
  {year} {2007})}\BibitemShut {NoStop}%
\bibitem [{\citenamefont {Avidor}\ and\ \citenamefont
  {Allison}(2016)}]{avidor2016helium}%
  \BibitemOpen
  \bibfield  {author} {\bibinfo {author} {\bibfnamefont {N.}~\bibnamefont
  {Avidor}}\ and\ \bibinfo {author} {\bibfnamefont {W.}~\bibnamefont
  {Allison}},\ }\bibfield  {title} {\bibinfo {title} {Helium diffraction as a
  probe of structure and proton order on model ice surfaces},\ }\href@noop {}
  {\bibfield  {journal} {\bibinfo  {journal} {The Journal of Physical Chemistry
  Letters}\ }\textbf {\bibinfo {volume} {7}},\ \bibinfo {pages} {4520}
  (\bibinfo {year} {2016})}\BibitemShut {NoStop}%
\bibitem [{\citenamefont {Tamt{\"o}gl}\ \emph {et~al.}(2020)\citenamefont
  {Tamt{\"o}gl}, \citenamefont {Sacchi}, \citenamefont {Avidor}, \citenamefont
  {Calvo-Almaz{\'a}n}, \citenamefont {Townsend}, \citenamefont {Bremholm},
  \citenamefont {Hofmann}, \citenamefont {Ellis},\ and\ \citenamefont
  {Allison}}]{tamtogl2020nanoscopic}%
  \BibitemOpen
  \bibfield  {author} {\bibinfo {author} {\bibfnamefont {A.}~\bibnamefont
  {Tamt{\"o}gl}}, \bibinfo {author} {\bibfnamefont {M.}~\bibnamefont {Sacchi}},
  \bibinfo {author} {\bibfnamefont {N.}~\bibnamefont {Avidor}}, \bibinfo
  {author} {\bibfnamefont {I.}~\bibnamefont {Calvo-Almaz{\'a}n}}, \bibinfo
  {author} {\bibfnamefont {P.~S.}\ \bibnamefont {Townsend}}, \bibinfo {author}
  {\bibfnamefont {M.}~\bibnamefont {Bremholm}}, \bibinfo {author}
  {\bibfnamefont {P.}~\bibnamefont {Hofmann}}, \bibinfo {author} {\bibfnamefont
  {J.}~\bibnamefont {Ellis}},\ and\ \bibinfo {author} {\bibfnamefont
  {W.}~\bibnamefont {Allison}},\ }\bibfield  {title} {\bibinfo {title}
  {Nanoscopic diffusion of water on a topological insulator},\ }\href@noop {}
  {\bibfield  {journal} {\bibinfo  {journal} {Nature Communications}\ }\textbf
  {\bibinfo {volume} {11}},\ \bibinfo {pages} {1} (\bibinfo {year}
  {2020})}\BibitemShut {NoStop}%
\bibitem [{\citenamefont {Chudley}\ and\ \citenamefont
  {Elliott}(1961)}]{chudley1961neutron}%
  \BibitemOpen
  \bibfield  {author} {\bibinfo {author} {\bibfnamefont {C.}~\bibnamefont
  {Chudley}}\ and\ \bibinfo {author} {\bibfnamefont {R.}~\bibnamefont
  {Elliott}},\ }\bibfield  {title} {\bibinfo {title} {Neutron scattering from a
  liquid on a jump diffusion model},\ }\href@noop {} {\bibfield  {journal}
  {\bibinfo  {journal} {Proceedings of the Physical Society}\ }\textbf
  {\bibinfo {volume} {77}},\ \bibinfo {pages} {353} (\bibinfo {year}
  {1961})}\BibitemShut {NoStop}%
\bibitem [{\citenamefont {Tuddenham}\ \emph {et~al.}(2010)\citenamefont
  {Tuddenham}, \citenamefont {Hedgeland}, \citenamefont {Jardine},
  \citenamefont {Lechner}, \citenamefont {Hinch},\ and\ \citenamefont
  {Allison}}]{tuddenham2010lineshapes}%
  \BibitemOpen
  \bibfield  {author} {\bibinfo {author} {\bibfnamefont {F.~E.}\ \bibnamefont
  {Tuddenham}}, \bibinfo {author} {\bibfnamefont {H.}~\bibnamefont
  {Hedgeland}}, \bibinfo {author} {\bibfnamefont {A.~P.}\ \bibnamefont
  {Jardine}}, \bibinfo {author} {\bibfnamefont {B.~A.}\ \bibnamefont
  {Lechner}}, \bibinfo {author} {\bibfnamefont {B.}~\bibnamefont {Hinch}},\
  and\ \bibinfo {author} {\bibfnamefont {W.}~\bibnamefont {Allison}},\
  }\bibfield  {title} {\bibinfo {title} {Lineshapes in quasi-elastic scattering
  from species hopping between non-equivalent surface sites},\ }\href@noop {}
  {\bibfield  {journal} {\bibinfo  {journal} {Surface science}\ }\textbf
  {\bibinfo {volume} {604}},\ \bibinfo {pages} {1459} (\bibinfo {year}
  {2010})}\BibitemShut {NoStop}%
\bibitem [{\citenamefont {Townsend}\ and\ \citenamefont
  {Avidor}(2019)}]{townsend2019signatures}%
  \BibitemOpen
  \bibfield  {author} {\bibinfo {author} {\bibfnamefont {P.~S.}\ \bibnamefont
  {Townsend}}\ and\ \bibinfo {author} {\bibfnamefont {N.}~\bibnamefont
  {Avidor}},\ }\bibfield  {title} {\bibinfo {title} {Signatures of multiple
  jumps in surface diffusion on honeycomb surfaces},\ }\href@noop {} {\bibfield
   {journal} {\bibinfo  {journal} {Physical Review B}\ }\textbf {\bibinfo
  {volume} {99}},\ \bibinfo {pages} {115419} (\bibinfo {year}
  {2019})}\BibitemShut {NoStop}%
\bibitem [{\citenamefont {Kole}\ \emph {et~al.}(2012)\citenamefont {Kole},
  \citenamefont {Hedgeland}, \citenamefont {Jardine}, \citenamefont {Allison},
  \citenamefont {Ellis},\ and\ \citenamefont
  {Alexandrowicz}}]{kole2012probing}%
  \BibitemOpen
  \bibfield  {author} {\bibinfo {author} {\bibfnamefont {P.~R.}\ \bibnamefont
  {Kole}}, \bibinfo {author} {\bibfnamefont {H.}~\bibnamefont {Hedgeland}},
  \bibinfo {author} {\bibfnamefont {A.~P.}\ \bibnamefont {Jardine}}, \bibinfo
  {author} {\bibfnamefont {W.}~\bibnamefont {Allison}}, \bibinfo {author}
  {\bibfnamefont {J.}~\bibnamefont {Ellis}},\ and\ \bibinfo {author}
  {\bibfnamefont {G.}~\bibnamefont {Alexandrowicz}},\ }\bibfield  {title}
  {\bibinfo {title} {Probing the non-pairwise interactions between co molecules
  moving on a cu (111) surface},\ }\href@noop {} {\bibfield  {journal}
  {\bibinfo  {journal} {Journal of Physics: Condensed Matter}\ }\textbf
  {\bibinfo {volume} {24}},\ \bibinfo {pages} {104016} (\bibinfo {year}
  {2012})}\BibitemShut {NoStop}%
\end{thebibliography}%

\end{document}